\documentclass[a4paper,fleqn,usenatbib]{mnras}
\usepackage[T1]{fontenc}
\usepackage{ae,aecompl}
\usepackage{graphicx}	
\usepackage{amsmath}	
\usepackage{amssymb}	
\renewcommand{\vec}[1]{\mbox{\boldmath $#1$}}
\def\lsim{\lower.4ex\hbox{$\;\buildrel <\over{\scriptstyle\sim}\;$}}
\title[Why do old stars not obey gyrochronology?] 
{How supercritical are stellar dynamos, or why do old main-sequence dwarfs not obey gyrochronology?}
\author[L.~Kitchatinov \& A.~Nepomnyashchikh]
{Leonid~Kitchatinov$^{1,2}$\thanks{E-mail: kit@iszf.irk.ru}
and Alexander~Nepomnyashchikh$^{1}$
\\
$^{1}$Institute for Solar-Terrestrial Physics, Lermontov Str. 126A, 664033, Irkutsk, Russia\\
$^{2}$Pulkovo Astronomical Observatory, St. Petersburg, 196140, Russia
}

\date{Accepted XXX. Received YYY; in original form ZZZ}

\pubyear{2017}
\begin{document}
\label{firstpage}
\pagerange{\pageref{firstpage}--\pageref{lastpage}}
\maketitle

\begin{abstract}
Asteroseismological determinations of stellar ages have shown that old main-sequence dwarfs do not obey gyrochronology. Their rotation is slow compared to young stars but faster than gyrochronology predicts. This can be explained by the presence of a maximum rotation period beyond which the large-scale dynamo switches off and stops providing global magnetic fields necessary for stellar spindown. Assuming this explanation, the excess of stellar dynamo parameters over their marginal values can be estimated for given spectral type and rotation rate. The estimation gives the dynamo number for the Sun about 10\% above its critical value. The corresponding dynamo model provides - though with some further tuning - reasonable results for the Sun. Following the same approach, the differential rotation and marginal dynamo modes are computed for stars between 0.7 and 1.2 solar masses. With increasing stellar mass, the differential rotation and the ratio of toroidal-to-poloidal field are predicted to increase while the field topology changes from dipolar to mixed quadrupolar-dipolar parity.
\end{abstract}


\section{Introduction}

Self-sustained hydromagnetic dynamos in general and stellar dynamos in particular can be understood as instabilities of conducting fluids to magnetic disturbances \citep[cf., e.g.,][]{M78}. The dynamo instability amplifies a pre-existing seed field and sustains it against Ohmic decay. Similar to all known instabilities, dynamos are active only if a certain controlling parameter exceeds a definite critical value. In the case of stellar dynamos, a properly normalized rotation rate can be chosen as the controlling parameter.

\citet{DL78} were probably the first to argue that stellar dynamos can produce global magnetic fields only if the ratio $\mathrm{Ro} = P_\mathrm{rot}/\tau$ of the rotation period $P_\mathrm{rot}$ to convective turnover time $\tau$ is not too large, $\mathrm{Ro} \lsim 1$. They explained the relative fast rotation of stars of spectral type earlier than F6 by their inability to sustain a global dynamo in their shallow convection zones. Dwarf stars of later spectral types are spinning down because of the angular momentum loss due to a magnetically coupled wind. Large-scale magnetic fields of presumably dynamo origin enhance the spin-down by increasing the effective radius of stellar wind emanation \citep{K67}. The rotation period of the dynamo-hosting stars increases with age $t$ according to the \citet{S72} law: $P_\mathrm{rot} \propto \sqrt{t}$. Empirical specification of the proportionality coefficient in the \citeauthor{S72} law as a function of stellar mass (or equivalent parameter) gave rise to gyrochronology~-- a determination of dwarf star ages from their rotation rates and mass \citep{B03,B07,Cea09,MMS09}.

Gyrochronology was considered to be an almost universal tool for determination of the ages of dwarf stars until the recent accumulation of asteroseismic data revealed its limitations. The data on stellar oscillations can be used in combination with photometric and spectroscopic data and stellar structure models to determine stellar ages \citep{Cea14,SS16}. \citet{Aea15} used the wealth of asteroseismic data from the {\it Kepler} mission to calibrate gyrochronology and found that there is no common gyrochronology relation for stars of all ages. Dwarf stars, which have passed more than about a half of their main-sequence life, do not obey gyrochronology. They rotate faster than gyrochronology predicts \citep{Sea16,Mea16}. The old stars rotate much slower than young stars of comparable mass but still faster than it follows from gyrochronology calibrated for young clusters, their magnetic activity being low.

From the standpoint of dynamo theory, the explanation for the deviation from gyrochronology  seems to be straightforward: the stars spin down to the marginal rotation rate for the onset of global dynamos where spindown stops or becomes inefficient. Though the explanation seems plausible, uncertainties in dynamo theory do not allow its confirmation by modelling stellar dynamos. This paper tries to progress in another direction: by reducing uncertainties in dynamo modelling on the basis of stellar rotation data.

Dynamo theory for stellar activity enjoyed a certain recent progress \citep[cf., e.g., review by][]{C17}. Parameters of dynamo models, nevertheless, cannot be confidently estimated. In particular, the value of the dynamo number and degree of the excess of the number over its critical value for the onset of dynamo instability are prescribed arbitrarily. This paper suggests the estimation of the maximum rotation period for dynamo operation as the function of $B-V$ colour. The estimation suggests the dynamo number for the Sun to be about 10\% above its critical value. The estimation is then used in a particular version of the flux-transport dynamo model pioneered by \citet{CSD95} and \citet{D95} to simulate the solar activity cycle. The simulations show - though with some further tuning - close correspondence to solar observations. The paper proceeds by modelling differential rotation and marginal dynamo regimes for a range of spectral types from late F to mid K. The computations predict the differential rotation and magnetic field topology for solar-type stars close to breakup of their global dynamos. An observational test of the predictions can further constrain dynamo models.
\section{Basic assumptions and method}\label{Method}
Dynamo models produce large-scale fields of finite amplitude for dynamo numbers $D > D_\mathrm{c}$ exceeding its critical value $D_\mathrm{c}$, and zero fields otherwise. We therefore assume that gyrochronology calibrated with relatively young stars is valid up to the age when the critical rotation period $P_\mathrm{c}$ is reached, beyond which the global dynamo is switched-off. The empirical relation by \citet{B07},
\begin{equation}
    P_\mathrm{rot} = a t^n\left(B - V\ -\ 0.4\right)^b\ \mathrm{d},
    \label{1}
\end{equation}
with $a = 0.773$, $n = 0.519$, and $b = 0.601$ is used to specify the rotation period of relatively young stars of given $B - V$ colour and age $t$ measured in Myr.

The maximum rotation period terminating validity of Eq.\,(\ref{1}) can be inferred from stellar rotation data. Figure\,1 of \citep{R84} clearly shows that rotation periods of solar-type stars do not exceed a maximum value dependent on $B-V$. \citeauthor{R84} approximated this maximum value by the linear relation, which can be written as
\begin{equation}
    P_\mathrm{c} = 111.3 \left( B - V\ - 0.405\right)\ \mathrm{d} .
    \label{2}
\end{equation}
He supposed that this approximation gives the rotation period for the end of the main-sequence life of a star. A combination of Eqs\,(\ref{1}) and (\ref{2}) can, however, show that the maximum rotation period is reached
at the age of
\begin{equation}
    t_\mathrm{c} \simeq 14.4 \left( B - V - 0.4\right)^{0.77}\
    \mathrm{Gyr} ,
    \label{t_c}
\end{equation}
roughly in the middle of main-sequence evolution \citep[in agreement with][]{Mea16}. In particular, for the solar value of $B - V = 0.656$, the maximum rotation period of about 28\,d follows for the age of about 5~Gyr. We suggest the dynamo-initiated spindown up to the maximum rotation period of Eq.\,(\ref{2}) where the dynamo stops as the approximate -- possibly not very precise -- scenario.

It may be noted that the expressions in brackets of Eqs~(\ref{1}) and (\ref{2}) are very similar implying that the stars of $B-V \lsim 0.4$ are not able to sustain large-scale dynamos in their shallow convection zones.

The main controlling parameter for the $\alpha\Omega$-dynamos which are believed to operate in the Sun and not too young solar-type stars is the dynamo number
\begin{equation}
    D = \frac{\alpha\Delta\Omega R^3}{\eta_{_\mathrm{T}}^2} ,
    \label{3}
\end{equation}
where $\eta_{_\mathrm{T}}$ is the eddy diffusivity, $R$ is the stellar radius, $\Delta\Omega$ is the differential rotation, and $\alpha$ is the parameter of the $\alpha$-effect of toroidal-to-poloidal field conversion \citep{KR80}. Which kind of $\alpha$-effect is more important -- the Babcock-Leighton (BL) mechanism \citep{B61} or the canonical effect of turbulent convection by \citet{P55} -- seems to be the main uncertainty of solar/stellar dynamo modeling. There is observational evidence for the operation of the BL mechanism on the Sun \citep{Dea10,KO11}. The  participation of the canonical $\alpha$-effect is favoured by theoretical arguments \citep{Pea14,Hea14}. Whatever mechanism produces the $\alpha$-effect, there is a general consensus about its origin from the action of the Coriolis force on relatively small-scale motions. The $\alpha$-parameter is therefore proportional to the rotation rate and can be written as
\begin{equation}
    \alpha = \alpha_\mathrm{c}P_\mathrm{c}/P_\mathrm{rot} ,
    \label{4}
\end{equation}
where $\alpha_\mathrm{c}$ is the critical $\alpha$-value of a particular dynamo model and $P_\mathrm{c}$ is the maximum rotation period of Eq.\,(\ref{2}). Sidereal rotation period $P_\mathrm{rot} =25.4$\,d and Eq.\,(\ref{2}) give the ratio $P_\mathrm{c}/P_\mathrm{rot} = 1.1$ for the Sun. The solar dynamo is, therefore, estimated to be about 10\% supercritical.

Other parameters in the dynamo number (\ref{3}) depend on rotation rate less than $\alpha$. The radius and the eddy diffusivity vary little with main-sequence age. The rapidly rotating solar twin AB Doradus is observed to possess nearly the same differential rotation as the Sun \citep{DC97}. Differential rotation modelling for intermediate rotation rates show only its moderate variation \citep{KO12}. The expression for dynamo number,
$D = D_\mathrm{c}P_\mathrm{c}/P_\mathrm{rot}$, similar to Eq.\,(\ref{4}), can therefore be suggested as a recipe for estimating the number in models for stars of given rotation period and colour. The recipe will be used below in a solar dynamo model, which seems to produce reasonable results.

The model is then applied to stars of different mass to estimate critical dynamo parameters and marginal dynamo modes. This is done as follows.
Structure of a star has to be specified to model its differential rotation and dynamo. The {\sl EZ} code by \citet{P04} is used to compute the evolutionary sequence of structure models for a star of given mass and metallicity $Z = 0.02$. For each evolutionary stage, the table of colour-temperature relations and the interpolation code of \citet{VBC03} are used to estimate the $B-V$ colour and gyrochronological rotation period of Eq.\,(\ref{1}). When the rotation period becomes close to the maximum rotation period of Eq.\,(\ref{2}), the star is supposed to be at the evolutionary stage of marginal dynamo. The corresponding structure model is then used to compute the differential rotation and global magnetic field modes of the marginal dynamo as described in Section \ref{model}.

The computations were performed for the mass in the range of $(0.7 - 1.2)M_\odot$ with step of $0.05M_\odot$. For smaller mass, the age of approaching the maximum period of Eq.\,(\ref{2}) exceeds the age of the Galaxy. For larger mass, modelling of dynamos in shallow convection zones becomes problematic. This mass range corresponds to $B-V$ from about 0.5 to 1.2. The empirical upper bound on the rotation period of Eq.\,(\ref{2}) was found by \citep{R84} for the range of $0.5 < B-V < 1$. We assume that Eq.\,(\ref{2}) remains valid beyond this range and extrapolate it up to $B-V = 1.2$. It may be noted, that stellar mass is not a convenient structure parameter. The structure depends on the chemical composition of a star and varies considerably with metallicity for given mass. The dependencies on mass and metallicity combine, however, into a common dependence on the effective temperature (or colour) when the temperature (colour) is used as the structure parameter \citep[cf., e.g., fig.\,1 in][]{KO12}. The $B-V$ colour is, therefore, used as the structure parameter.
\section{Dynamo model}\label{model}
The lack of data on the internal magnetic field of the Sun precludes a selection of a conventional  dynamo model from those hitherto proposed. The flux-transport (advection-dominated) models with BL mechanism of poloidal field re-generation as the $\alpha$-effect \citep{CSD95,D95}  provide, however, the closest agreement with surface observations. The model of this paper belongs to this class of dynamo model. It is almost identical to the model of a previous publication \citep[][KN17 hereafter]{KN17} where more details on the model design can be found.

The model solves the mean-field induction equation,
\begin{equation}
    \frac{\partial{\vec B}}{\partial t} =
    {\vec\nabla}\times\left({\vec V}{\times}{\vec B}\ +\ {\vec{\cal E}} \right),
    \label{5}
\end{equation}
in a spherical layer of a stellar convection zone. In this equation, $\vec{\cal E}$ is the mean electromotive force \citep[EMF, cf.][]{KR80} which includes the effects of turbulent diffusion, diamagnetic pumping and the $\alpha$-effect specified below. The usual spherical coordinates $(r,\theta,\phi)$ are used and axial symmetry of the large-scale magnetic ($\vec B$) and velocity ($\vec V$) fields about the rotation axis is assumed:
\begin{eqnarray}
{\vec B} &=& {\vec e}_\phi B + {\vec\nabla}\times
\left({\vec e}_\phi\frac{A}{r\sin\theta}\right)\ ,
\nonumber \\
{\vec V} &=& {\vec e}_\phi r\sin\theta\Omega + \frac{1}{\rho}{\vec\nabla}\times
\left({\vec e}_\phi\frac{\psi}{r\sin\theta}\right)\ ,
    \label{6}
\end{eqnarray}
where $B$ is the toroidal field, $A$ is the poloidal field potential, $\Omega$ is the angular velocity, $\psi$ is the stream function of the meridional flow, and ${\vec e}_\phi$ is the azimuthal unit vector.

The flow $\vec V$ is specified using the differential rotation model of  \citet{KO11DR,KO12}. A peculiarity of the model is that the eddy transport coefficients are not prescribed but expressed in terms of the entropy gradient, the entropy being one of the dependent variables of the model. The eddy viscosity, in particular, is specified as
\begin{equation}
    \nu_{_\mathrm{T}} = -\frac{\tau\ell^2 g}{15c_\mathrm{p}}
    \frac{\partial S}{\partial r} ,
    \label{7}
\end{equation}
where $S$ is the entropy, $g$ is gravity, $\tau$ and $\ell$ are the convective temporal and spatial scales, respectively, and $c_\mathrm{p}$ is the specific heat at constant pressure. The only significant modification of the model since 2011 is that the mixing length $\ell$ is reduced near the base of the convection zone. This length is usually assumed to be proportional to the local pressure scale height: $\ell_0 = \alpha_{_\mathrm{MLT}}H_\mathrm{p}$. It seems, however, plausible that the length should decrease on approaching the the base below which the convection does not penetrate:
\begin{equation}
    \ell = \ell_\mathrm{min} +\frac{1}{2}\left(\ell_0 - \ell_\mathrm{min}\right)
    \left[ 1 + \mathrm{erf}\left(\frac{r/r_\mathrm{i} - x_\ell}{d}\right)\right] ,
    \label{8}
\end{equation}
where $\mathrm{erf}$ is the error function, $r_\mathrm{i}$ is the radius of the base of the convection zone, $\ell_\mathrm{min} = 0.01R$, $x_\ell = 1.01$, and $d = 0.025$. The model with such a re-defined length-scale (\ref{8}) was used to specify the differential rotation, meridional flow and the eddy diffusivities for dynamo simulations.

The azimuthal component of Eq.\,(\ref{5}) gives the dynamo equation for the toroidal field:
\begin{eqnarray}
    \frac{\partial B}{\partial t} &=&
    \frac{1}{r}\left(\frac{\partial\Omega}{\partial r}\frac{\partial A}{\partial\theta} -
    \frac{\partial\Omega}{\partial\theta}\frac{\partial A}{\partial r}\right)
    \nonumber \\
    &+& \frac{1}{\rho r^2}\frac{\partial\psi}{\partial r}
    \frac{\partial}{\partial\theta}\left( \frac{B}{\sin\theta} \right)
    -\frac{1}{r\sin\theta}\frac{\partial\psi}{\partial\theta}
    \frac{\partial}{\partial r} \left(\frac{B}{\rho r} \right)
    \nonumber \\
    &+& \frac{1}{r}\left(\frac{\partial (r{\cal E}_\theta )}{\partial r} -
    \frac{\partial {\cal E}_r}{\partial\theta}\right) .
    \label{9}
\end{eqnarray}
In this equation, the first line on the right-hand side describes the toroidal field production by the differential rotation, the second line - the field advection by the meridional flow, and the third line - the turbulent transport.

Our model includes the anisotropy of turbulent transport induced by rotation. The rather complicated expression for EMF for rotating fluids \citep{P08} can be split in three relatively simple parts representing physically different effects of turbulent diffusion (${\vec{\cal E}}^\mathrm{diff}$), diamagnetic pumping (${\vec{\cal E}}^\mathrm{dia}$), and the $\alpha$-effect (${\vec{\cal E}}^\alpha$):
\begin{equation}
    {\vec{\cal E}} = {\vec{\cal E}}^\mathrm{diff}
    + {\vec{\cal E}}^\mathrm{dia}
    + {\vec{\cal E}}^\alpha .
    \label{10}
\end{equation}
The diffusive part of the EMF reads
\begin{equation}
    {\vec{\cal E}}^\mathrm{diff} = -\eta{\vec\nabla}\times{\vec B}
    - \eta_\|{\vec e}\times\left({\vec e}\cdot{\vec\nabla}\right){\vec B} ,
    \label{11}
\end{equation}
where $\vec e$ is the unit vector along the rotation axis, $\eta$ is the isotropic part of the eddy diffusivity, and $\eta_\|$ is the additional diffusivity for the direction along the rotation axis. The rotationally induced anisotropy results in a difference between the coefficients of eddy diffusion for the directions along ($\eta + \eta_\|$) and across ($\eta$) the rotation axis,
\begin{equation}
    \eta = \eta_{_\mathrm{T}}\phi(\Omega^*), \ \
    \eta_\| = \eta_{_\mathrm{T}}\phi_\|(\Omega^*) .
    \label{12}
\end{equation}
In this equation, $\eta_{_\mathrm{T}}$ is the diffusivity for non-rotating fluid and the dependence on rotation rate enters via the functions $\phi(\Omega^*)$ and $\phi_\|(\Omega^*)$ of the Coriolis number
\begin{equation}
    \Omega^* = 2\tau\Omega .
    \label{13}
\end{equation}
In the slow rotation limit, $\Omega^* \rightarrow 0$, $\phi \rightarrow 1$, $\phi_\| \rightarrow 0$, and Eq.\,(\ref{11}) reduces to the usual expression for isotropic diffusion.

Diamagnetic pumping is important for solar and presumably for stellar dynamos \citep{KKT06,GG08}. The part of EMF responsible for the pumping effect reads
\begin{eqnarray}
    {\vec{\cal E}}^\mathrm{dia} &=& -({\vec\nabla}\tilde{\eta})\times{\vec B} + ({\vec\nabla}\eta_\|)\times{\vec e}({\vec e}\cdot{\vec B}),
    \nonumber \\
    \tilde{\eta} &=& \eta_{_\mathrm{T}}\phi_1(\Omega^*)\ ,
    \label{14}
\end{eqnarray}
where the second term on the right-hand side is again caused by the rotational anisotropy. The references for the explicit expressions for the functions $\phi$, $\phi_\|$, and $\phi_1$ are given in KN17. The near-bottom sub-adiabatic layer of penetrative convection with reduced diffusion is important for the effect of diamagnetic pumping. The non-local effect of penetration is not accounted for by the local estimation (\ref{7}) for the eddy viscosity. To account for the penetration layer, we reduce the magnetic eddy diffusivity in a thin layer near the bottom similar to the Eq.\,(\ref{8}) for the mixing length:
\begin{equation}
    \eta_{_\mathrm{T}} = \frac{1}{\mathrm{Pm}}\left[ \nu_\mathrm{i} +
    \frac{1}{2}(\nu_{_\mathrm{T}} - \nu_\mathrm{i})
    \left(1 + \mathrm{erf}\left(\frac{r/r_\mathrm{i} - x_\eta}{d}\right)\right)\right]\ ,
    \label{15}
\end{equation}
where $\nu_{_\mathrm{T}}$ is the eddy viscosity of Eq.\,(\ref{7}), $\mathrm{Pm}$ is the magnetic Prandtl number, $\nu_\mathrm{i} = 10^{-4}\times\nu_{_\mathrm{T}}^\mathrm{max}$ ($\nu_{_\mathrm{T}}^\mathrm{max}$ is the maximum value of $\nu_{_\mathrm{T}}$ within the convection zone), and $x_\eta = 1.1$.

Only the azimuthal component of $\vec{\cal E}^\alpha$ contributes to the $\alpha\Omega$-dynamo equations. We prescribe the ${\cal E}_\phi^\alpha$ as follows
\begin{equation}
    {\cal E}^\alpha_\phi = \alpha\frac{B(r_\mathrm{i},\theta)}{1 + (B(r_\mathrm{i},\theta)/B_0)^2} F(\theta)\phi_\alpha(r/r_\mathrm{e})\ ,
    \label{16}
\end{equation}
where $r_\mathrm{e}$ is the radius of the external boundary of the simulation domain ($r_\mathrm{e} = 0.97R$ in all computations of this paper), the functions $F(\theta)$ and $\phi_\alpha (x)$ define the distributions of the $\alpha$-effect over latitude and radius, respectively. $B_0 = 10$ \,kG slightly above the energy equipartition value for deep solar convection was used in the solar dynamo model. Computations of the marginal modes of stellar dynamos do not need a specification of $B_0$.
The BL mechanism is related to finite tilts of the surface bipolar active regions \citep{B61}. The poloidal field generation is, therefore prescribed to occur near the surface and at relatively low latitudes:
\begin{eqnarray}
    \phi_\alpha (x) &=& \frac{1}{2}\left[ 1 + \mathrm{erf}\left( (x + 2.5h_\alpha -1)/h_\alpha\right)\right] ,
    \nonumber \\
    F(\theta) &=& \cos\theta\sin^{n_\alpha}\theta .
    \label{17}
\end{eqnarray}
$h_\alpha = 0.02$ in the computations of this paper. The $\alpha$-effect of Eq.\,(\ref{16}) is non-local in space: the poloidal field is generated near the surface from the bottom toroidal field $B(r_\mathrm{i},\theta)$.

The poloidal field equation reads
\begin{equation}
    \frac{\partial A}{\partial t} = \frac{1}{\rho r^2\sin\theta}
    \left( \frac{\partial\psi}{\partial r}\frac{\partial A}{\partial\theta}
    - \frac{\partial\psi}{\partial\theta}\frac{\partial A}{\partial r} \right)
    + r\sin\theta\ {\cal E}_\phi .
    \label{18}
\end{equation}
The explicit expressions for the EMF components ${\cal E}_r$, ${\cal E}_\theta$ and ${\cal E}_\phi$ in the dynamo equations (\ref{9}) and (\ref{18}) can be found from Eqs.\,(\ref{11}), (\ref{14}) and (\ref{16}). We do not write these explicit expressions in the dynamo equations to avoid complexity. The boundary conditions correspond to the interface with a superconductor at the inner boundary,
\begin{equation}
    {\cal E}_\theta = 0,\ \ A = 0\ \ \mathrm{at}\ r = r_\mathrm{i},
    \label{19}
\end{equation}
and vertical field at the external boundary,
\begin{equation}
    B = 0,\ \ \frac{\partial A}{\partial r} = 0,\ \ \mathrm{at}\  r = r_\mathrm{e} .
    \label{20}
\end{equation}

The best fit to the 11-year duration of the solar cycle and dipolar parity of magnetic fields is obtained with the Prandtl number $\mathrm{Pm} = 3$ in Eq.\,(\ref{15}) and $n_\alpha = 7$ in Eq.(\ref{17}) (KN17). All computations of this paper were performed with these parameter values. The method of numerical solution of the dynamo equations is described in KN17. The model does not normalize variables to dimensionless units and the dimensional $\alpha$-parameter of Eq.\,(\ref{16}) is the governing parameter of the model.
\section{Results}
\subsection{Test case: the Sun}
Estimations of Section\,\ref{Method} suggest that the $\alpha$-parameter for the Sun exceeds its critical value for onset of the dynamo-instability by about 10\%. We proceed by estimating the critical value and constructing a nonlinear model for the solar dynamo with the 10\% super-criticality to evaluate the performance of the above-described approach and to illustrate its main steps before applying it to other stars.

\begin{figure}
	\includegraphics[width=\columnwidth]{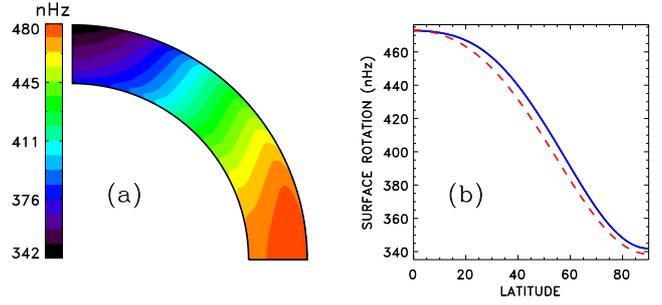}
    \caption{Angular velocity iso-contours in the convection zone (a) and the
        surface rotation profile (b) of the solar rotation model. Doppler measurements of \citet{SU90} are shown by the dashed line in the right panel for comparison.
        }
    \label{f1}
\end{figure}

The angular velocity distribution, the meridional flow and diffusivity profiles are specified with the differential rotation model. The differential rotation of Fig.\,\ref{f1} is close to the results of helioseismology \citep{WBL97,Sea98}.

\begin{figure}
	\includegraphics[width=\columnwidth]{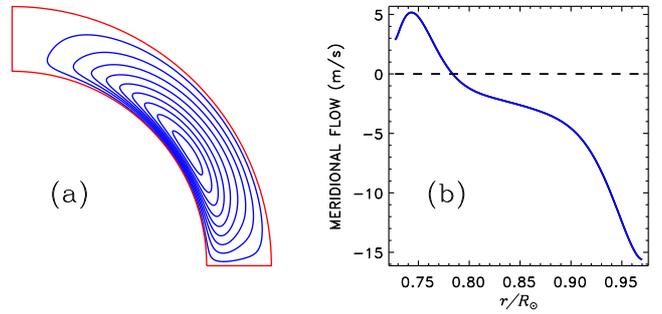}
    \caption{Stream lines of the meridional flow (a) and depth profile of the
        flow velocity for $45\degr$ latitude (b). Positive velocity means the flow towards the equator.
        }
    \label{f2}
\end{figure}

Our model does not include a tachocline. The detections of the convection zone inner boundary at $r_\mathrm{i} = 0.713R_\odot$ by \citet{CGT91} and \citet{BA97} and measurements of the tachocline central radius and thickness by \citet{Cea99} place the entire tachocline beneath the convection zone. Moreover, only radial rotational shear is large in the tachocline that requires an appreciable radial component of the magnetic field for generating the toroidal field. The redial field should be, however, small near the base of the convection zone. Otherwise it is a fossil field penetrating from the radiative interior. The fossil field is not included in our dynamo model.

The modelled meridional flow is shown in Fig.\,\ref{f2}. The flow consists of a single circulation cell with a poleward surface flow and return equator-ward flow of some meters per second near the bottom. The decline of the flow with depth near the base is caused by a decrease in the characteristic scale (\ref{8}). The slow circulation avoids a confident helioseismic measurement. Fig.\,\ref{f2} is however quite similar to recent results of \citet{RA15}.

\begin{figure}
	\includegraphics[width=8 truecm]{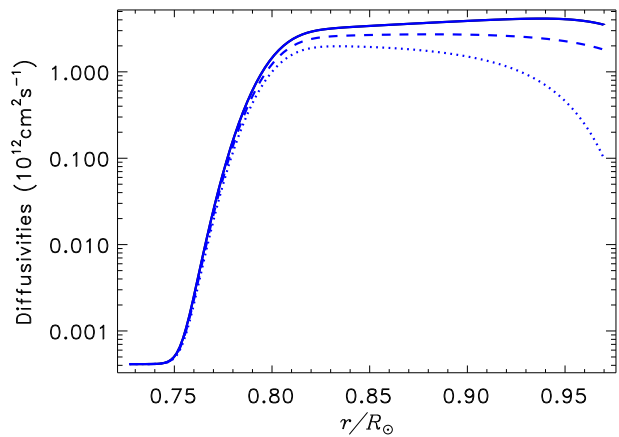}
    \caption{Depth profiles of the diffusion coefficients $\eta$ ({\sl full line}),
        $\tilde\eta$ ({\sl dashed}) and $\eta_\|$ ({\sl dotted}) of Eqs\,(\ref{11}) and (\ref{14}) estimated from the entropy gradient supplied by the differential rotation model.
        }
    \label{f3}
\end{figure}

An important \lq side product' of the differential rotation model is the entropy distribution. The distribution can be used for evaluation of the eddy diffusion with Eqs~(\ref{7}) and (\ref{15}). So-defined profiles of the eddy diffusivities are shown in Fig.\,\ref{f3} and used in the dynamo model. The entropy varies little with latitude. The diffusion coefficients are, therefore, assumed to depend on radius only.

The results of Figs \ref{f1} to \ref{f3} provide sufficient input for dynamo modelling. The modelling gives the critical value $\alpha_\mathrm{c}^\mathrm{d} = 0.158$\,m\,s$^{-1}$ for the onset of the dynamo (a small difference in this value with KN17 is caused by the difference in the parameterization of Eq.\,(\ref{8}) for the mixing length). The upper index \lq$\mathrm{d}$' in the notation for the critical $\alpha$ means that it corresponds to the field of dipolar (equator-antisymmetric) parity. The quadrupolar fields require higher $\alpha_\mathrm{c}^\mathrm{q} = 0.169$\,m\,s$^{-1}$ for excitation. The smaller value of $\alpha_\mathrm{c}^\mathrm{d}$ implies that nonlinear simulations starting from an initial field of mixed parity eventually approach the dipolar parity.

\begin{figure}
	\includegraphics[width=\columnwidth]{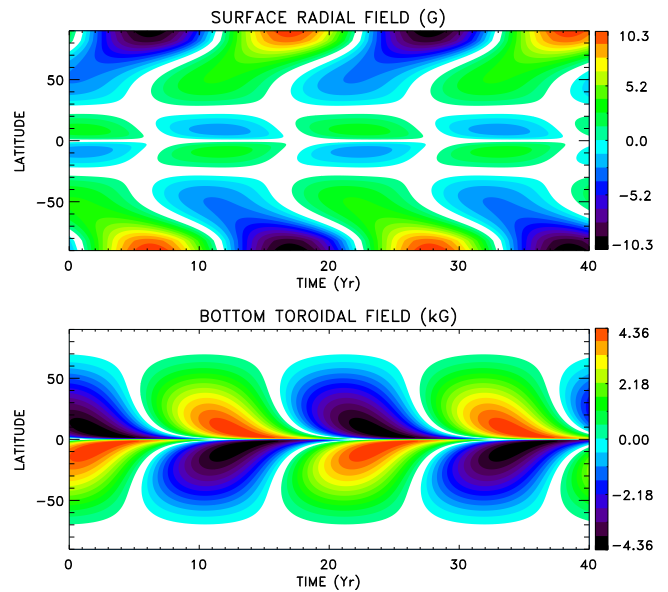}
    \caption{Time-latitude diagrams of magnetic fields for the model of the solar dynamo
        computed with 10\% supercritical $\alpha = 0.174$\,m\,s$^{-1}$.
        }
    \label{f4}
\end{figure}

Figure~\ref{f4} illustrates such an asymptotic regime for the model of the solar dynamo computed with about 10\% supercritical value of $\alpha = 0.174$\,m\,s$^{-1}$. The equatorial drift of the toroidal field in Fig.\,\ref{f4} is caused by the deep meridional flow \citep{HKC14}. The polar drift of the surface poloidal field is a result of the common action of diffusion and advection. It has to be noted that the values of the magnetic Prandtl number $\mathrm{Pm} = 3$ and $n_\alpha = 7$ in Eq.\,(\ref{17}) were chosen to fit the observed period of the solar cycle and dipolar parity of surface fields (KN17). Correspondence of the model of Fig.\,\ref{f4} to solar observations, however, is not limited to the parity and cycle duration. The model is based on the estimation of the excess of the $\alpha$-parameter over its marginal value described in the Section\,\ref{Method}. We proceed by computing marginal modes of stellar dynamos.

\begin{figure}
	\includegraphics[width=\columnwidth]{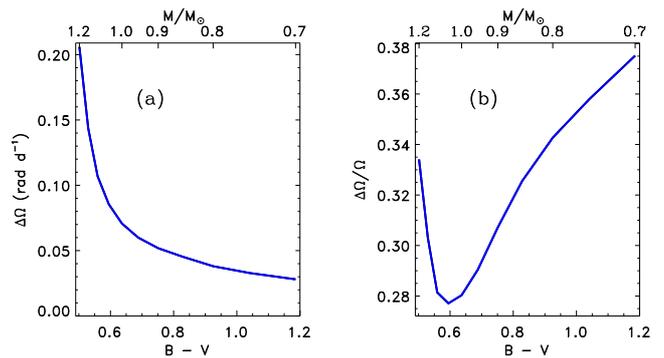}
    \caption{Surface difference in angular velocities
        between the equator and poles (a) and its normalized
        value (b) depending on $B-V$ colour of stars hosting marginal large-scale dynamos. The corresponding scale of fractional mass is shown on the top.
        }
    \label{f5}
\end{figure}
\subsection{Marginal stellar dynamos}
Differential rotation is an important input parameter of dynamo models. The differential rotation computed for the stars, which are supposed to be in a state of marginal dynamos, are shown in Fig.\,\ref{f5} as a function of $B-V$ colour. The relation of $B-V$ to the rotation period $P_\mathrm{c}$ or age $t_\mathrm{c}$ of these stars is given by equation (\ref{2}) or (\ref{t_c}), respectively.
The range of $B-V$ in this figure corresponds to mature dwarfs in the mass range $(0.7 - 1.2)M_\odot$ and metallicity $Z = 0.02$.

The increasing trend of the differential rotation with surface temperature was observed by \citet{Bea05} for young rapid rotators. Left panel of Fig.\,\ref{f5} shows the slow rotation counterpart of this trend. The surface differential rotation increases strongly with stellar mass in the mass range of our computations. It may be noted, however, that the stars of smaller mass are older, and they rotate slower, according to  Eqs\,(\ref{2}) and (\ref{t_c}). The relative value of the differential rotation $\sim 30\%$ in the right-hand panel of this Figure varies moderately with $B-V$.

In contrast to the differential rotation, two other inputs for the dynamo model supplied by the computations of differential rotation - the deep meridional flow and eddy diffusivity - are not observable. We, therefore, mention only that the diffusivity and near-bottom meridional velocity generally increase with decreasing $B-V$. The diffusivity $\eta$ estimated for the middle of convection zone increases from about $1.2\times10^8$\,m$^2$s$^{-1}$ for $B-V = 1.19$ ($0.7M_\odot$) to about $5.2\times10^8$\,m$^2$s$^{-1}$ for $B-V = 0.529$ ($1.15M_\odot$). The corresponding increase in near-bottom meridional velocity at $45\degr$ latitude is from 1.9 to 3.2\,m\,s$^{-1}$. Both quantities then decrease slightly to the smallest $B-V = 0.502$ ($1.2M_\odot$) of our computations.

\begin{figure}
	\includegraphics[width=7 truecm]{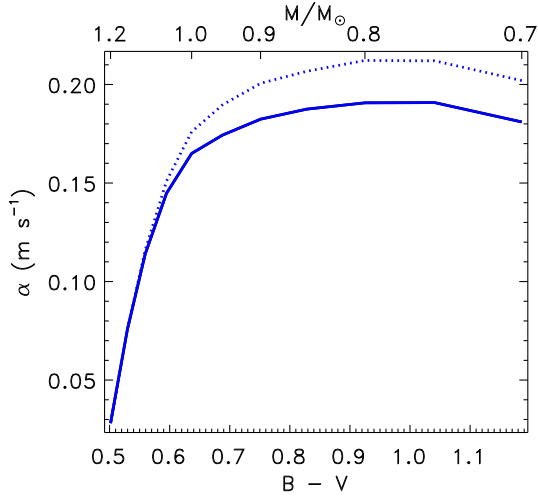}
    \caption{Marginal values of $\alpha$ for the onset of the dynamo computed for critical rotation periods of Eq.\,(\ref{2}). The two lines correspond to the magnetic field modes of dipolar ({\sl full line}) and quadrupolar ({\sl dotted}) parity.
        }
    \label{f6}
\end{figure}

Figure\,\ref{f6} shows the marginal values of the $\alpha$-parameter for generation of the global fields of dipolar and quadrupolar parity. Similar to the Sun, $\alpha^\mathrm{d}_\mathrm{c} < \alpha^\mathrm{q}_\mathrm{c}$ for stars of relatively low mass. The slowly rotating low-mass stars are, therefore, predicted to host dipolar global fields. The critical $\alpha$-values for both parities are practically equal on the low $B-V$ side of Fig.\,\ref{f6}. The late F-stars are predicted to possess less regular fields of mixed parity. The preference of a certain parity requires some physical agent to link N and S hemispheres for their coherent operation in the generation of the magnetic field \citep{CNC04,HY10}. The link becomes inefficient in shallow convection envelopes of F stars, at least in the dynamo model of this paper. It may be also noted that F-stars of our sample are younger, and they rotate faster compared to cooler stars. Zeeman-Doppler imaging of \citet{Sea_ZDI_16} have shown that complexity of the field topology increases for smaller age or higher rotation rate.

The computations of marginal dynamos cannot define the field amplitude. The ratio of the toroidal-to-poloidal field strengths can, however, be estimated. The ratio of the amplitudes of bottom toroidal to surface poloidal (polar) fields increased with decreasing $B-V$ from about 340 ($B-V = 1.19$) to 1010 ($B-V = 0.50$). The simulated large-scale field is increasingly hidden inside the convection zones of stars of increasing mass.

\begin{figure}
	\includegraphics[width=\columnwidth]{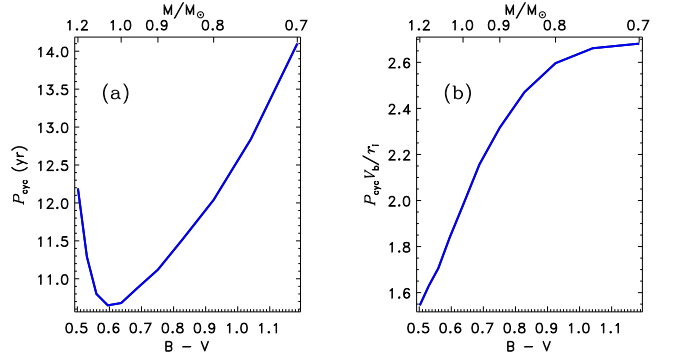}
    \caption{Cycle period for marginal dynamos (a) and the period ratio
        to the advection time (b). $V_\mathrm{b}$ is the meridional velocity at the inner boundary $r_\mathrm{i}$ at the latitude of $45\degr$.
        }
    \label{f7}
\end{figure}

All the simulated dynamos were cyclic. The cycle periods are shown in Fig.\,\ref{f7}. The long cycles from 10 to 14 yr of this figure are typical of old slow rotators \citep{SB99}. The cycle duration is believed to be controlled by the meridional flow. Fig.\,\ref{f7} shows also the ratio of the cycle period to advection time. The ratio -- though not constant -- is of the order one and varies moderately.

\begin{figure}
	\includegraphics[width=\columnwidth]{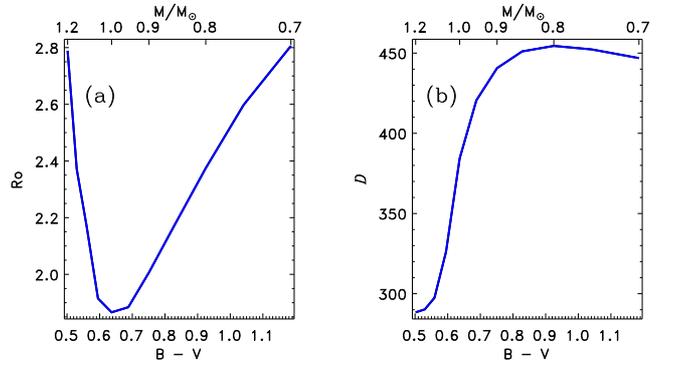}
    \caption{(a) The Rossby number estimated for the rotation periods
        of Eq.\,(\ref{2}). (b) The marginal dynamo number (\ref{3}) estimated with the isotropic part $\eta$ of the eddy diffusivity at middle depth in the convection zone.
        }
    \label{f8}
\end{figure}

Fig.\,\ref{f8} shows estimations of the Rossby and dynamo numbers. The Rossby number is position-dependent. The estimations of Fig.\,\ref{f8} were done for the radius $r$ one pressure scale hight above the base of the convection zone, i.e. for the radius $r$ where $r - r_\mathrm{i} = H_\mathrm{p}$. The Rossby number $\mathrm{Ro} \sim 2$ of Fig.\,\ref{f8} agrees with estimation of \citet{Sea16}.
The two numbers in Fig.\,\ref{f8} are often used to parameterize observations or theoretical computations of magnetic activity. The two parameters of this figure vary moderately but are not constant. This implies that stellar activity can be only crudely parameterized by a single parameter.
\section{Conclusions}
Asteroseismic measurements of stellar ages revealed a dramatic decrease in the rate of stellar spindown and magnetic activity for ages beyond about a half of the main-sequence life \citep{Sea16,Mea16}. This paper suggests an explanation for the violation of gyrochronology by a switch-off of the global dynamo for a rotation period exceeding the critical value estimated by Eq.\,(\ref{2}). The switch-off should not be abrupt. Dynamo theory explains the observed variations in amplitudes and periods of solar cycles by fluctuations in dynamo parameters \citep[cf.][]{C11}. The fluctuations will cause a star to wander between the parameter regions of (supercritical) dynamo of \lq normal' magnetic activity and (subcritical) dynamo of fading activity if the rotation period is close to its critical value. The dynamo models with fluctuating parameters reproduce the statistics of grand solar minima \citep{USK07} only if the dynamo number is close to its marginal value \citep{Mea08,OK13,KKB15}. It may be appropriate to note that observations of total solar eclipses in the Maunder minimum did not find an extended solar corona \citep{E76} indicating a relatively small co-rotation radius and inefficient spindown.

Unfortunately, the concept of dynamo switch-off at the critical rotation period cannot be supported by dynamo modelling because of uncertainties in key dynamo parameters. The uncertainties can, however, be reduced using this concept. Eq.\,(\ref{4}) for $\alpha$-parameter or similar equation for the dynamo number can be used to estimate the dynamo parameter values for a star of given rotation period and colour. The estimation gives the excess of about 10\% over the critical value for the Sun and computation with the 10\% super-criticality provides a reasonable model for the solar dynamo. The concept was then used to estimate the activity cycle periods and global field structure for stars of different mass rotating with periods close to the critical value of Eq.\,(\ref{2}). The estimations predict a gradual change from a dipolar surface field in K-stars to more disordered mixed parity fields in late F-stars.

It may be noted that the estimation of Eq.\,(\ref{4}) is not limited to the particular dynamo model of this paper. The critical value of the dynamo number is model-dependent but the recipe for estimating an appropriate (supercritical) value for a star of given $P_\mathrm{rot}$ and $B-V$ is not.

We finally note that Fig.\,\ref{f8} indicates that one parameter -- the Rossby number or dynamo number -- cannot fully characterize the operation of dynamo, at least, for our model.
\citet{Gea12} found that magnetic field topology of late-type stars  depends strongly on stellar structure. In particular, relative magnitude of toroidal and poloidal field components differ between the fully convective stars and stars with appreciable radiative cores. The structural changes can hardly be accounted for with one parameter --  rotation rate or Rossby number.
One parameter, therefore, can parameterize stellar magnetic activity only crudely. It seems, however, plausible that the dynamo mechanism is uniquely defined by the rotation rate and structure of a star. Two parameters -- say, rotation rate and colour -- should, therefore, define uniquely long-term stellar activity. Two is not many.
\section*{Acknowledgements}
This work was supported by the Russian Foundation for Basic Research (project 17-02-00016).
\bibliographystyle{mnras}

\bsp	
\label{lastpage}
\end{document}